# LHC LUMINOSITY UPGRADE WITH LARGE PIWINSKI ANGLE SCHEME: A RECENT LOOK*

C. M. Bhat,[#] Fermilab, Batavia, IL 60510, U.S.A. and
F. Zimmermann, CERN, Geneva, Switzerland

*Abstract*
Luminosity upgrade at the LHC collider using longitudinally flat bunches in combination with the large crossing angle (large Piwinski angle scheme) is being studied with renewed interest in recent years. By design, the total beam-beam tune shift at the LHC is less than 0.015 for two interaction points together. But the 2010-11 3.5 TeV collider operation and dedicated studies indicated that the beam-beam tune shift is >0.015 per interaction point. In view of this development we have revisited the requirements for the Large Piwinski Angle scheme at the LHC. In this paper we present a new set of parameters and luminosity calculations for the desired upgrade by investigating 1) current performance of the LHC injectors, 2) e-cloud issues on nearly flat bunches and 3) realistic beam particle distributions from longitudinal beam dynamics simulations. We also make some remarks on the needed upgrades on the LHC injector accelerators.

## INTRODUCTION

Significant theoretical progress has been made with regards to the choice of the bunch length and crossing angle near the beam-beam limit at the LHC collider in order to maximize luminosity [1, 2]. It has been shown that [1, 3] by using super-bunches in circular machines like the Large Hadron Collider (LHC) one can achieve pp collider luminosities $>10^{35}$cm$^{-2}$sec$^{-1}$. However, the collider detectors at the interaction points demand as small a number of interactions per bunch crossing as possible. For example, at the luminosity $\sim 10^{34}$cm$^{-2}$sec$^{-1}$ the allowed number of interactions per collision is about 20. With super bunches or very long bunches the number of interactions per collision is likely to go up significantly. Consequently one can foresee to use shorter flat bunches of total length $\sim$ 0.5 meter for LHC scenario and at the same time take advantage of luminosity that would come from the long flat bunches. The peak luminosity at the interaction point for two colliding round Gaussian beams of equal bunch intensity, with rms bunch length $\sigma_z$ (<< $\beta^*$) and transverse size of $\sigma^*_{trans}$ is given by,

$$L = \frac{n_b f_{rev} N_p^2}{4\pi \sigma^{*2}_{trans}} F \cong \frac{n_b f_{rev} N_p^2}{4\pi \sigma^{*2}_{trans}} \times \frac{1}{\sqrt{1 + \left[\frac{\theta_c \sigma_z}{2\sigma^*_{trans}}\right]^2}} \quad (1)$$

where $n_b$, $f_{rev}$ $N_p$ and $\theta_c$, are number of bunches, revolution frequency, number of protons per bunch and full crossing angle of the two beams at the interaction point, respectively. The quantity $\theta_c \sigma_z / 2\sigma^*_{trans}$ is known as the "Piwinski angle". However, it has been shown that the total incoherent beam-beam tune shift for colliding beams is also reduced by a similar factor $F$ [4], for LHC like collider which has crossing angle in alternating horizontal-vertical planes. Then, the Eq. (1) in terms of beam-beam tune shift $\Delta Q_{bb}(\cong N_b r_p F/2\pi\varepsilon_n)$ becomes [4],

$$L = \gamma \Delta Q_{bb}^2 \frac{\pi n_b f_{rev} \varepsilon_n}{r_p^2 \beta^*} \frac{1}{F} \quad (2)$$

with $r_p$, $\varepsilon_n$ and $\beta^*$ are classical radius of the proton, average normalized transverse emittance of the colliding beams and lattice function at the interaction point in the ring, respectively. Consequently, at a fixed value of $\Delta Q_{bb}$, it is possible to increase the luminosity by increasing Piwinski angle, i.e., increasing the crossing angle and/or bunch length along with bunch intensity.

Recently a levelled peak luminosity goal of about $5\times 10^{34}$cm$^{-2}$sec$^{-1}$ had been defined for the LHC upgrade scenarios [5] in order to keep the peak number of interactions per collisions at about 200. Further, the 2010-11 collider operation and dedicated beam-beam studies indicated that the beam-beam tune shift attainable in operation is higher than the design assumption and more than 0.015 per interaction point has been achieved with good luminosity lifetime [6]. These changes have prompted us to revisit the parameter list particularly for the large Piwinski angle (LPA) scheme.

## REALISTIC BUNCH PROFILES

In reality the bunch profiles in the LHC are not Gaussian. The measured bunch profile in the buckets of 400 MHz rf system is more like a Hoffman-Pedersen distribution. To create nearly flat bunches one can use a combination of 400 MHz and 800 MHz rf (which is being designed) or of 200 MHz (cavities exist but not installed in the LHC) and 400 MHz rf systems in bunch lengthening mode (BLM). Recent studies in the CERN PS with the combination of h=21 and 42 showed that the beam in the BLM is more stable against coupled bunch instability [7], while in the SPS beam studies with h=4620 and 18480 showed that the beam becomes less stable in BLM [8]. Nevertheless, it is well established that to stabilize the beam in the SPS one needs the combination of h=4620 and 18480 with bunch shortening mode (BSM). So, it is foreseen to add a second harmonic rf system in the LHC to stabilize the high intensity beam for future operation. In any of the above cases the bunch profile will significantly deviate from the Gaussian shape.

---

Table 1: New parameter list for the LPA scheme at the LHC at 7 TeV. "BLMpt5" and "BSMpt5" represent the rf voltage ratio of 0.5 with 180 deg phase apart for 1st and 2nd harmonic rf cavities.

| Parameters | | Nominal | Ultimate | LPA (200MHz+400MHz RF) BLMpt5 (A) | LPA (200MHz+400MHz RF) BSMpt5 (B) | LPA (400MHz+800MHz RF) BLMpt5 (C) | LPA (400MHz+800MHz RF) BSMpt5 (D) |
|---|---|---|---|---|---|---|---|
| Number of Bunches | | 2808 | 2808 | 1404 | 1404 | 1404 | 1404 |
| Protons/bunch | $N_b(10^{11})$ | 1.15 | 1.7 | 3.9 | 3.3 | 3.5 | 3.1 |
| Beam Current [A] | | 0.58 | 0.86 | 1 | 0.84 | 0.88 | 0.78 |
| Norm. Transv. Emit | um | 3.75 | 3.75 | 3.0-3.75 | 3.0-3.75 | 3.0-3.75 | 3.0-3.75 |
| $\sigma z$ | cm | 7.55 | 7.55 | 16 | 11 | 9 | 6 |
| Bunch Spacing | nsec | 25 | 25 | 50 | 50 | 50 | 50 |
| $\beta^*$ at IP1 and IP5 | m | 0.55 | 0.5 | 0.25 | 0.25 | 0.36 | 0.36 |
| $\theta_c$ | urad | 285 | 315 | 380 | 380 | 380 | 380 |
| Piwinski Angle | | 0.64 | 0.75 | 3.03-2.71 | 2.08-1.86 | 1.42-1.22 | 0.963-0.862 |
| $\Delta Q_{bb}$ | | 0.006 | 0.009 | 0.01-0.009 | 0.012-0.01 | 0.016-0.014 | 0.018-0.015 |
| Peak and Average Lum. (10 hr turn around) | $10^{34}$cm$^{-2}$s$^{-1}$ | 1 / 0.46 | 2.3 / 0.91 | 6-5.3 / 1.68-1.6 | 5.9-5.2 / 1.5-1.4 | 6.1-5.3 / 1.6 - 1.5 | 6.0-5.0 / 1.5 -1.4 |
| Event Pileup | | 19 | 44 | 201-227 | 224-196 | 232-200 | 227-192 |

In view of these we have performed beam dynamics simulations of the LHC beam profiles for BLM and BSM modes starting from the Hoffman-Pedersen distribution generating final distributions which are then used for our luminosity estimations. Figure 1 shows a simulated bunch profile for a combination of 400 MHz and 800 MHz rf waveforms.

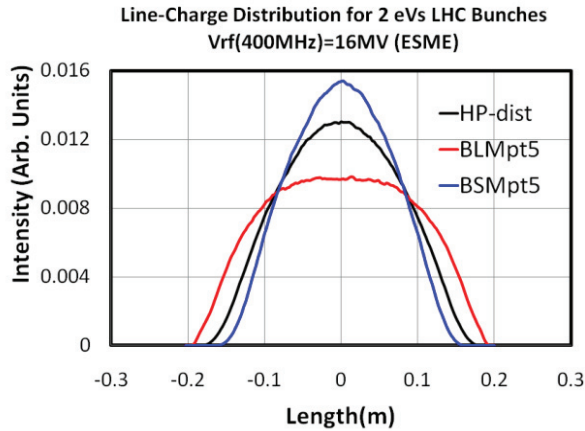

Figure 1: Simulated line-charge distributions for various combinations of 1st and 2nd harmonic of the main rf system in the LHC.

Table 1 displays a revised parameter list for the LHC luminosity upgrade considering four different cases as shown. We assumed Vrf(200 MHz)= 3 MV for cases "A" and "B", and Vrf(400 MHz)=16 MV for cases "C" and "D". In both cases we have investigated the BLMpt5 and BSMpt5 cases. Crossing angle, beam emittance and bunch spacing were assumed to be 380 μrad, 2 eVs and a 50 nsec, respectively.

We find that for the same total beam current as in the case of "ultimate luminosity" scenario, we can gain nearly a factor of two in luminosity. Further, the BSM modes require about 15% less beam current as compared with the BLM mode. No doubt, the gain in luminosity is coming from the increase in bunch intensity [5].

The maximum bunch intensities that the LPA scheme demands are substantially larger than those needed for nominal or ultimate LHC requirements. In view of this, we summarize the current status of the injector accelerators and the required improvements.

## CURRENT PERFORMANCE OF THE LHC INJECTORS

Over the last decade several improvements/upgrades have been undertaken at the LHC injector accelerators to

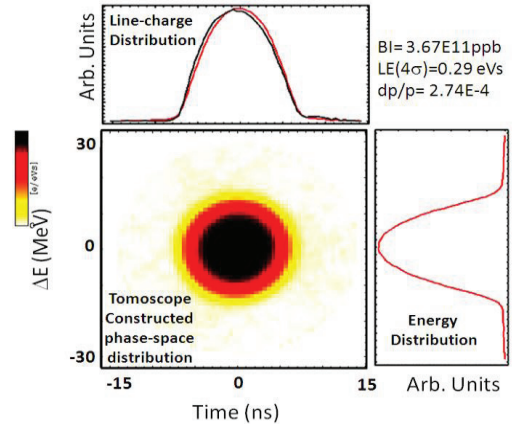

Figure 2: Tomoscope reconstruction of the phase-space distribution of beam particles in a single bunch in the PS just before extraction at 26 GeV. The line-charge distribution is measured using a wall current monitor and the energy distribution is generated from the distribution.

provide the beam required to reach the "nominal" ($10^{34}$cm$^{-2}$sec$^{-1}$) or "ultimate" LHC luminosities

($2.3\times10^{34}$cm$^{-2}$sec$^{-1}$). The LHC injector chain consists of four accelerators – LINAC (50 MeV), PSB (1.4 GeV), PS (26 GeV) and SPS (450 GeV). The full potential of the injectors is being investigated. The PS in combination with the PSB will generate the bunches needed for various LHC filling patterns. Currently, 36 bunches (with <$1.7\times10^{11}$ppb) of 50 ns bunch spacing (or 72 bunches of 25 ns on demand) are routinely transferred from the PS to the SPS for the LHC collider operation.

LPA requirement of a single bunch intensity of ~$3.6\times10^{11}$ppb with longitudinal emittance LE($4\sigma$)~0.3 eVs and transverse emittance ~$3\mu$m is reached [9] in the PS and accelerated with >90% efficiency in the SPS. A tomoscope reconstruction of the phase space distribution of particles [10] in a single bunch of nearly highest intensity at 26 GeV in the PS is illustrated in Fig. 1. The results so far are very promising. However, significant improvement in rf power and beam loading compensation are needed to achieve the PS performance necessary to fulfil the LPA requirements for multiple high intensity bunches.

The SPS offers real challenges for high intensity beam acceleration from 26 GeV to 450 GeV. One of the major problems is e-cloud induced beam instability. The issues related to beam intensity limits in the SPS prior to 2010 are reviewed in [11]. Recent studies [12] in the SPS showed that the transverse mode coupling instability threshold is at about $1.6\times10^{11}$ppb with the nominal lattice and close to zero chromaticity, but a new optics with lower transition energy [13] and/or larger chromaticity (Q'~2) allow acceleration of bunches with significantly higher intensity (>$3\times10^{11}$ppb).

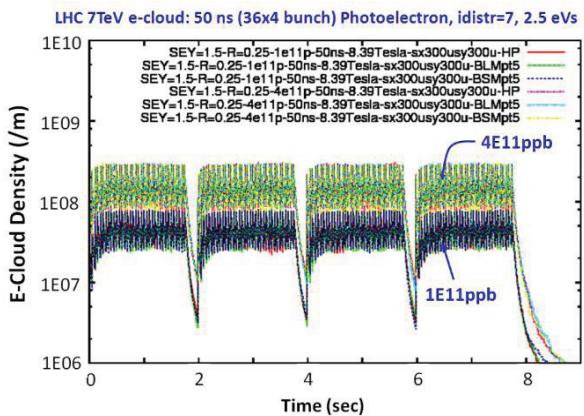

Figure 3: E-cloud simulations for realistic bunch profiles.

## E-CLOUD ISSUES

Recent beam studies [14] have shown that the e-cloud related vacuum activities and beam instability are next in line to be addressed for the high intensity operation of the LHC. Detailed e-cloud simulations for all upgrade scenarios were performed in the past both for standard Gaussian bunches as well as for ideal flat bunches [5,15].

In view of the new LPA parameter list shown in Table 1, we have re-evaluated the e-cloud generation for bunches shown in Fig. 1 using an improved version of ECLOUD. Preliminary results show that the difference in e-cloud build-up for three different bunch profiles at same bunch intensity is negligible. Figure 3 shows simulation results for 4 batches of 36 bunch each with 50 nsec bunch spacing. A more detailed study of the e-cloud effect based on the recent measurements is in progress.

In conclusion we have presented an updated parameter list for the LPA scheme of the LHC luminosity upgrade based on the recent operational experience of the LHC complex.

The work is supported by Fermi Research Alliance and US LHC Accelerator Research Program (LARP) and by EuCARD-AccNet. One of the authors (CMB) would like to thank CERN for its hospitality and special thanks are due to O. Brüning, H. Damerau, E. Shaposhnikova, E. Mahner, and many CERN collaborators.